\documentclass[figures]{epl}

\newcommand\msg{p}
\newcommand\prob{p}
\newcommand\reg[1]{{\mbox{\scriptsize ${#1}$-reg}}}
\newcommand\cor[1]{{\mbox{\scriptsize ${#1}$-core}}}
\newcommand\eref[1]{(\ref{#1})}

\title{Sudden emergence of $q$-regular subgraphs in random graphs}
\shorttitle{$q$-regular subgraphs in random graphs}
\author{M. Pretti\inst{1} \and M. Weigt\inst{2}}

\institute
{
  \inst{1} INFM-CNR, Dipartimento di Fisica, Politecnico di
  Torino, Corso Duca degli Abruzzi~24, I-10129 Torino, Italy \\
  \inst{2} Institute for Scientific Interchange,
  Viale Settimio Severo~65, I-10133 Torino, Italy
}

\pacs{05.20.-y}{Classical statistical mechanics}

\pacs{02.50.-r}{Probability theory, stochastic processes and
statistics}

\pacs{89.20.-a}{Interdisciplinary applications of physics}

\begin{document}

\maketitle

\begin{abstract}
We investigate the computationally hard problem whether a random graph
of finite average vertex degree has an extensively large $q$-regular
subgraph, i.e., a subgraph with all vertices having degree equal
to~$q$. We reformulate this problem as a constraint-satisfaction
problem, and solve it using the cavity method of statistical physics
at zero temperature. For $q=3$, we find that the first large
$q$-regular subgraphs appear discontinuously at an average vertex
degree $c_\reg{3} \simeq 3.3546$ and contain immediately about $24\%$
of all vertices in the graph. This transition is extremely close to
(but different from) the well-known 3-core percolation point
$c_\cor{3} \simeq 3.3509$. For $q>3$, the $q$-regular subgraph
percolation threshold is found to coincide with that of the $q$-core.
\end{abstract}

\section{Introduction} In the last years, statistical
physics has increasingly been able to analyze and solve complex
problems coming from graph theory and theoretical computer
science~\cite{HaWe,PeIsMo}. The interest was particularly focused to
so-called {\it random constraint-satisfaction problems}, which are
characterized by a large number of discrete degrees of freedom being
subject to an also large number of hard constraints on subsets of
variables. The best-known examples are the satisfiability problem,
where a set of logical variables is asked to fulfil simultaneously a
large number of logical clauses, and the graph-coloring problems,
where vertices of a graph are to be assigned colors in a way that no
pair of neighboring vertices is equally colored. For both problems,
current mathematical tools in discrete mathematics, probability
theory, and theoretical computer science do not succeed in solving the
models completely.  Conversely, new approaches based on the
statistical mechanics of disordered systems, in particular the {\it
cavity method}~\cite{MezardParisi2001}, have crucially contributed to
our understanding, providing a framework to characterize the
statistical properties of the solution space of various
constraint-satisfaction problems, and to locate phase transitions in
its structure and
organization~\cite{MezardParisiZecchina2002,MuletPagnaniWeigtZecchina2002}.

In this letter, we address a graph-theoretical problem, which at a
first glance looks more related to percolation theory than to
constraint-satisfaction problems. The question is whether a random
graph of given finite connectivity possesses an extensively large
$q$-{regular subgraph}, i.e., a subgraph where every vertex has
exactly $q$ neighbors (constant degree~$q$). At a closer look,
this problem can be naturally embedded into the framework of
constraint-satisfaction problems and solved thereby using the
cavity method. On the other hand, typical tools from random graph
percolation theory are not able to solve the problem. The major
reason for this failure is that the problem is
NP-complete~\cite{GareyJohnson1979}, which means in particular
that no linear-time algorithm for searching $q$-regular subgraphs
exists. Mathematical tools based on the analysis of such
algorithms, in particular Wormald's rate equation
approach~\cite{Wormald1995}, are thus unavailable.  This is also
the major difference with respect to the apparently similar
problem of the existence of an extensively large $q$-{\it core},
i.e., the largest subgraph with degrees being equal to {\it or
larger} than~$q$. Such subgraph can be easily found by iteratively
removing all vertices of smaller degree. Using the rate equation
approach, Pittel, Spencer, and Wormald have
shown~\cite{PittelSpencerWormald1996}, that such a $q$-core
appears discontinuously at some average random graph degree
$c_\cor{q}$, and its size jumps from zero to a finite fraction of
all vertices. Let us notice that the existence of an extensive
$q$-core is a necessary condition for a giant $q$-regular subgraph
to exist, since each $q$-regular subgraph is by definition part of
the $q$-core. Such condition is by no means sufficient, i.e., a
$q$-core may in principle appear before $q$-regular subgraphs
exist at all, so that $c_\cor{q}$ is a lower bound to the
emergence of $q$-regular subgraphs. Bollobas, Kim, and
Verstra\"ete~\cite{BollobasKimVerstraete2006}, using a refined
version of the first-moment method (in statistical physics known
better as the {\it annealed approximation}), have proved that, for
$q=3$, there exists some gap $\gamma > 0$ such that, for $c \in
(c_\cor{q}, c_\cor{q} + \gamma)$, {\it almost surely} no
$q$-regular subgraph exists ({\it almost surely} means {\it with
probability tending to 1 in the thermodynamic limit} of infinitely
large graphs). Moreover, the authors conjecture that the same
holds true for any $q>3$. Looking a bit closer~\cite{PrettiWeigt}
to their proof, it is possible to determine the maximal~$\gamma$
compatible with the first-moment method, which turns out to be as
small as $\gamma \simeq 0.0003$ for $q=3$. This result means that
the currently best lower bound for the emergence of $3$-regular
subgraphs is $3.3512$, compared to $c_\cor{3} \simeq
3.3509$~\cite{PittelSpencerWormald1996}. Conversely, for $q>3$,
the first-moment method cannot improve the lower bound with
respect to $c_\cor{q}$, since no positive $\gamma$ is found.

Are the two transitions really so close to each other? How large
is the first $q$-regular subgraph to appear? How many of these
subgraphs are there, and in which way are they related to the
$q$-core? Is the above mentioned conjecture true? Such questions
have motivated us to address the problem from the point of view of
statistical physics. Using the cavity method, we have found
answers, which we argue to be exact.

\section{The model} Let us start with a more precise
definition of the problem. We study random graphs from the
Erd\"os-R\'enyi ensemble
$\mathcal{G}(N,c/N)$~\cite{ErdosRenyi1960}. They have $N$
vertices, and each pair of vertices is connected independently by
an edge with probability $c/N$. The scaling ${\cal O}(N^{-1})$
guarantees that the average vertex degree remains finite in the
thermodynamic limit $N\to \infty$, and tends to~$c$. The
probability distribution of the degree~$d$ approaches a Poissonian
of mean $c$, i.e., $\mathcal{P}_d=e^{-c} c^d/d!$. For this graph
ensemble, we ask in general for the existence of $q$-regular
subgraphs. More precisely, we ask whether there exists a {\it
threshold value} $c_\reg{q}$, such that, in the thermodynamic
limit, the probability of finding an extensively large $q$-regular
subgraph tends to $0$ for $c < c_\reg{q}$, and to $1$ for $c >
c_\reg{q}$. To answer this, we decorate each edge~$\{i,j\}$ with a
binary degree of freedom $x_{ij}=x_{ji}\in\{1,0\}$, meaning
respectively that the edge is, or is not, in a $q$-regular
subgraph. The constraints are associated to the vertices of the
original graph: In each vertex~$i$, either $0$ or $q$ ``active''
links can be present, i.e., $\sum_{j \in \partial i} x_{ij}
\in\{0,q\}$, where the sum runs over the set~$\partial i$ of all
neighbors of vertex~$i$.  Note that there is always a solution to
these constraints: $x_{ij} \equiv 0 \ \forall \{i,j\}$, i.e., the
empty subgraph.

\begin{figure}
  \onefigure[width=0.60\textwidth]{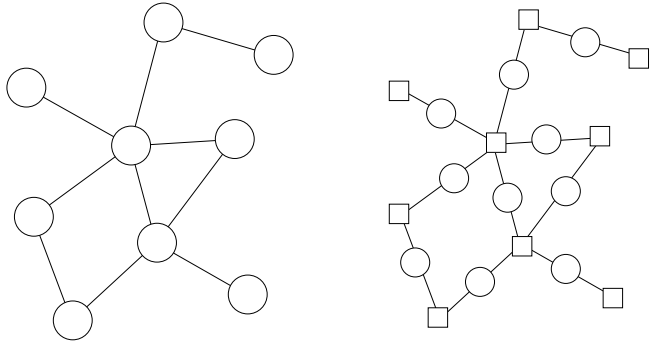}
  \caption
  {
    Original graph (left) and its factor graph representation (right).
    The vertices of the original graph become function nodes (squares),
    whereas the edges of the original graph become decorated by variable
    nodes (circles).
  }
  \label{fig:factorgraph}
\end{figure}
The bipartite structure of variables and constraints can be
represented by a suitable factor graph (see
fig.~\ref{fig:factorgraph}). Variable nodes are associated to the
edges of the original graph, and have thus constant degree~$2$ in
the factor graph, whereas function nodes are identified with the
original vertices, so that they have the Poissonian degree
distribution $\mathcal{P}_d$ of the original graph. We can assign
an energy cost to any global configuration $\{x_{ij}\}$ by means
of a Hamiltonian counting the number of violated constraints,
\begin{equation}
  \mathcal{H}(\{x_{ij}\})
  = \sum_{i=1}^N
  \biggr[ 1
  - \delta \biggr( \!\!
  \sum_{\ j \in \partial i} x_{ij} , 0 \biggr)
  - \delta \biggr( \!\!
  \sum_{\ j \in \partial i} x_{ij} , q
  \biggr)
  \biggr]
  \,,
  \label{eq:hamiltonian}
\end{equation}
with $\delta(\cdot, \cdot)$ denoting a Kronecker delta. Proper
$q$-regular subgraphs are zero-energy ground states of this
Hamiltonian, and their properties can be analyzed using the cavity
method at zero temperature. The main task is to calculate the
quenched entropy density
\begin{equation}
  s = \lim_{N \to \infty} N^{-1} \overline{\ln Z}
  \,,
  \label{eq:entropy1}
\end{equation}
where the zero-temperature partition function
\begin{equation}
  Z = \sum_{\{x_{ij}\}} \delta(\mathcal{H},0)
\end{equation}
represents the number of $q$-regular subgraphs,
while the overline denotes the average over the random graph
ensemble. The phase transition point~$c_\reg{q}$ can be identified
as the average vertex degree where this entropy first takes a
positive value.

Note that Hamiltonian (\ref{eq:hamiltonian}) contains two special
cases which were recently addressed with very similar methods.
First, for $q=1$, the problem of matchings in random graphs is
recovered, cf.~\cite{Zhou2003,Lenka}. Second, for $q=2$, the
regular subgraphs are
loops~\cite{MarinariMonasson2004,MarinariMonassonSemerjian2006,MaSe}.
There is, however, one big difference: For $q\leq 2$, the
existence problem becomes polynomially solvable, and the resulting
physical picture of the phase transition is simpler. In a random
graph, extensive matchings exist whenever there is an extensive
number of links ($c>0$), and extensive loops appear continuously
at the random-graph percolation transition~\cite{ErdosRenyi1960}
at $c=1$. The hard problem addressed in the afore-mentioned papers
is therefore the {\it counting problem}, whereas, for $q \geq 3$,
even asking about the very existence of $q$-regular subgraphs is
NP-complete.

\section{The replica-symmetric solution} The central step of
the cavity method is to set up a {\it message passing} procedure,
incorporating the local consequences of the degree constraints
into a globally self-consistent iterative scheme. We denote by
$\msg_{i\to j}$ an elementary message, i.e., the probability that
constraint~$i$ forces edge~$\{i,j\}$ to be present in the
subgraph. This happens if and only if exactly $q-1$ other edges
incident to~$i$ are present. We can therefore write
\begin{equation}
  \msg_{i \to j}
  =
  \frac
  {w_{\partial i \setminus j \, \to \, i}^{q-1}}
  {\sum_{n=0,q-1,q} w_{\partial i \setminus j \, \to \, i}^{n}}
  \,,
  \label{eq:cavity1}
\end{equation}
where we have defined
\begin{equation}
  w_{V \to i}^{n}
  \equiv
  \sum_{\stackrel{\scriptstyle U \subseteq V}{|U|=n}}
  \prod_{j \in U} \msg_{j \to i}
  \prod_{k \in V \setminus U} (1 - \msg_{k \to i})
  \,.
  \label{eq:cavity2}
\end{equation}
According to~\eref{eq:cavity2}, the numerator of
eq.~\eref{eq:cavity1} counts the joint probability that exactly
$q-1$ of the edges arriving in $i$ from other vertices than $j$
(i.e., from vertices in the set $\partial i \setminus j$) are
forced to be in the subgraph by their second end-vertex. The
denominator sums over all possibilities to have $0$, $q$, or $q-1$
such incoming edges. Note that only these three possibilities are
consistent with the constraint, i.e., normalization explicitly
excludes contradictory situations. Note also that the joint
probability is assumed to factorize in the edges, which, in the
replica-symmetric situation (one thermodynamic state
only~\cite{MezardParisi2001}), is expected to become exact for $N
\gg 1$, due to the locally tree-like organization of random
graphs. As a last remark, we note that $\msg_{i \to j} \equiv 0$
is always a solution of eqs.~(\ref{eq:cavity1},\ref{eq:cavity2}),
corresponding to the empty subgraph, which trivially satisfies all
degree constraints.

Having found a fixed point of these equations, we calculate
observables like the probabilities $p_i$, $p_{ij}$ that a vertex
or an edge, respectively, are in a $q$-regular subgraph,
\begin{eqnarray}
  \prob_{i}
  & = &
  \frac
  {w_{\partial i \to i}^{q}}
  {\sum_{n=0,q} w_{\partial i \to i}^{n}}
  \,,
  \label{eq:p-vertex}
  \\
  \prob_{ij}
  & = &
  \frac{\msg_{i \to j} \msg_{j \to i}}
  {\msg_{i \to j} \msg_{j \to i} + (1-\msg_{i \to j})(1-\msg_{j \to i})}
  \,.
  \label{eq:p-edge}
\end{eqnarray}
The numerator of eq.~\eref{eq:p-vertex} counts the probability
that $q$ edges arriving in vertex~$i$ are forced to be in the
subgraph, and it has to be normalized by the sum over all
consistent possibilities: Either $q$ edges (vertex in the
subgraph) or $0$ edges (vertex not in the subgraph) can be in the
subgraph. The numerator of eq.~\eref{eq:p-edge} contains the case
that the messages coming from the end-vertices of edge~$\{i,j\}$
consistently force the edge to be element of the subgraph, and it
has to be normalized with respect to all consistent message pairs.
Further on, we can calculate the entropy, which results from
eq.~\eref{eq:entropy1} via
\begin{equation}
  \ln Z
  =
  \sum_{i=1}^{N} \ln z_i
  - \sum_{\{i,j\}} \ln z_{ij}
  ,
  \label{eq:entropy2}
\end{equation}
where $z_i$ and $z_{ij}$ denote respectively the denominators of
the right hand sides of eqs.~\eref{eq:p-vertex}
and~\eref{eq:p-edge}.

Eqs.~(\ref{eq:cavity1},\ref{eq:cavity2}) can be solved either
directly on single graph instances, or in distribution in the
average over the random graphs. Due to the random structure of the
underlying graph, these equations are easily translated into a
self-consistent equation for the message
distribution~$\rho(\msg)$,
\begin{equation}
  \rho(\msg)
  =
  \sum_{d=0}^\infty e^{-c} \frac{c^d}{d!}
  \int_0^1 \upd \msg_1 \, \rho(\msg_1)
  \dots
  \int_0^1 \upd \msg_d \, \rho(\msg_d)
  \,
  \delta \left( \msg - f_d \left(
  \msg_1,\dots,\msg_d \right) \right)
  \,,
  \label{eq:message_distribution}
\end{equation}
where $\delta(\cdot)$ denotes a Dirac delta, and with $f_d$ being
given by eqs.~(\ref{eq:cavity1},\ref{eq:cavity2}). In complete
analogy, we can also derive equations for the ensemble-averaged
distribution of true occupation probabilities $p_i$, $p_{ij}$ and
for the entropy. The trivial solution $\rho(\msg)=\delta(\msg)$
exists independently of $c$, and has zero entropy. The question of
the existence of extensive $q$-regular subgraphs reduces to the
question of the existence and thermodynamic stability of
non-trivial solutions $\rho(\msg)$. A full solution can be
constructed only numerically, using a population-dynamical
scheme~\cite{MezardParisi2001}, but important information about
the onset of a non-trivial solution, and its relation to the
$q$-core, can be read off analytically from
eq.~(\ref{eq:message_distribution}). To do so, we first simplify
it by projecting the real-valued probabilities $\msg$ to a ternary
variable $X=0,1,*$, depending on whether $\msg=0,1$, or
$0<\msg<1$. The corresponding weights in $\rho(\msg)$ are denoted
by $P_X$. The trivial solution has of course $P_0 \equiv 1$ and
$P_1 \equiv P_* \equiv 0$, whereas in general we have to derive
closed equations for the three probabilities $P_X$, for $X=0,1,*$.
This task becomes considerably simplified by the observation that
$P_1$ must vanish, since, in the thermodynamic limit, a finite
fraction of messages polarized to $\msg=1$ would necessarily lead
to contradictions to the degree constraints. Moreover, it turns
out that a nontrivial ($X=*$) out-message is sent if and only if
$q-1$ or more in-messages are also nontrivial. This translates to
\begin{equation}
  P_{*}
  =
  \sum_{d=q-1}^{\infty}
  e^{-c} \frac{c^d}{d!}
  \sum_{n=q-1}^{d}
  {d \choose n} P_{*}^{n} P_{0}^{d-n}
  =
  e^{-cP_{*}}
  \sum_{n=q-1}^{\infty}
  \frac{(cP_{*})^n}{n!}
  ,
\end{equation}
where we have eliminated $P_0$ using normalization $P_0+P_*=1$.
This equation is not new: It appears in the context of the
$q$-core, and its first non-trivial solution $P_*>0$ appears
exactly at the $q$-core threshold~\cite{PittelSpencerWormald1996}.
Here, we find this connectivity as the {\it spinodal point} for
$q$-regular subgraphs: The first non-trivial solution of
eq.~(\ref{eq:message_distribution}) appears at this point, but its
thermodynamic stability is not guaranteed in principle. This
observation also clarifies the relation between the $q$-core and
all possible $q$-regular subgraphs: Edges not belonging to the
$q$-core never belong to any $q$-regular subgraph, whereas almost
all edges in the $q$-core belong to some but not all $q$-regular
subgraphs.

Unfortunately, the projection to ternary variables does not allow
for the calculation of the entropy, which is the important
thermodynamic potential for checking which solution of
eq.~(\ref{eq:message_distribution}) is actually the
thermodynamically stable one. As previously mentioned, the entropy
depends on the full information carried by $\rho(\msg)$, so that
we have to analyze eq.~(\ref{eq:message_distribution})
numerically. We have done so, using a representation of
$\rho(\msg)$ via a population of $2^{20}$~elements, and have used
a usual iterative update scheme~\cite{MezardParisi2001}.

\begin{figure}
  \resizebox{70mm}{!}{\includegraphics*[35mm,155mm][165mm,245mm]{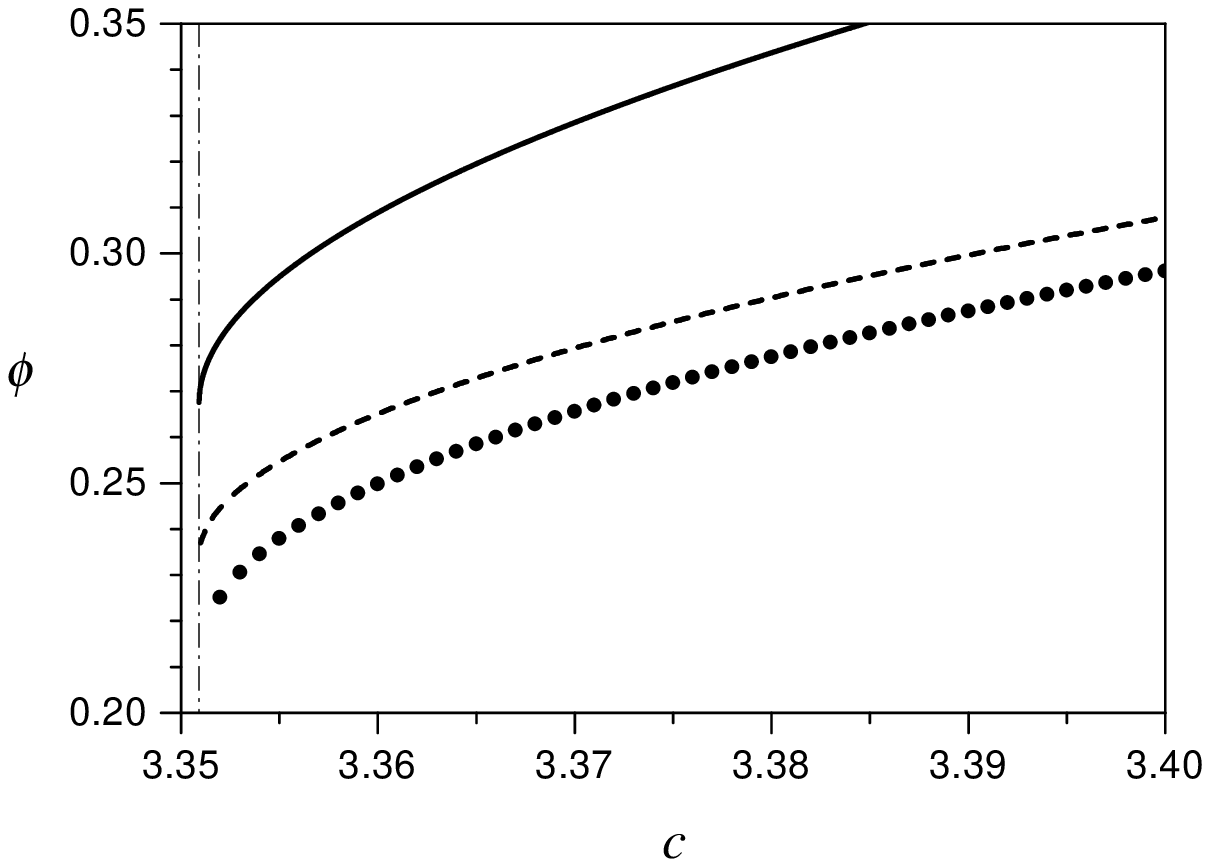}}
  \resizebox{70mm}{!}{\includegraphics*[35mm,155mm][165mm,245mm]{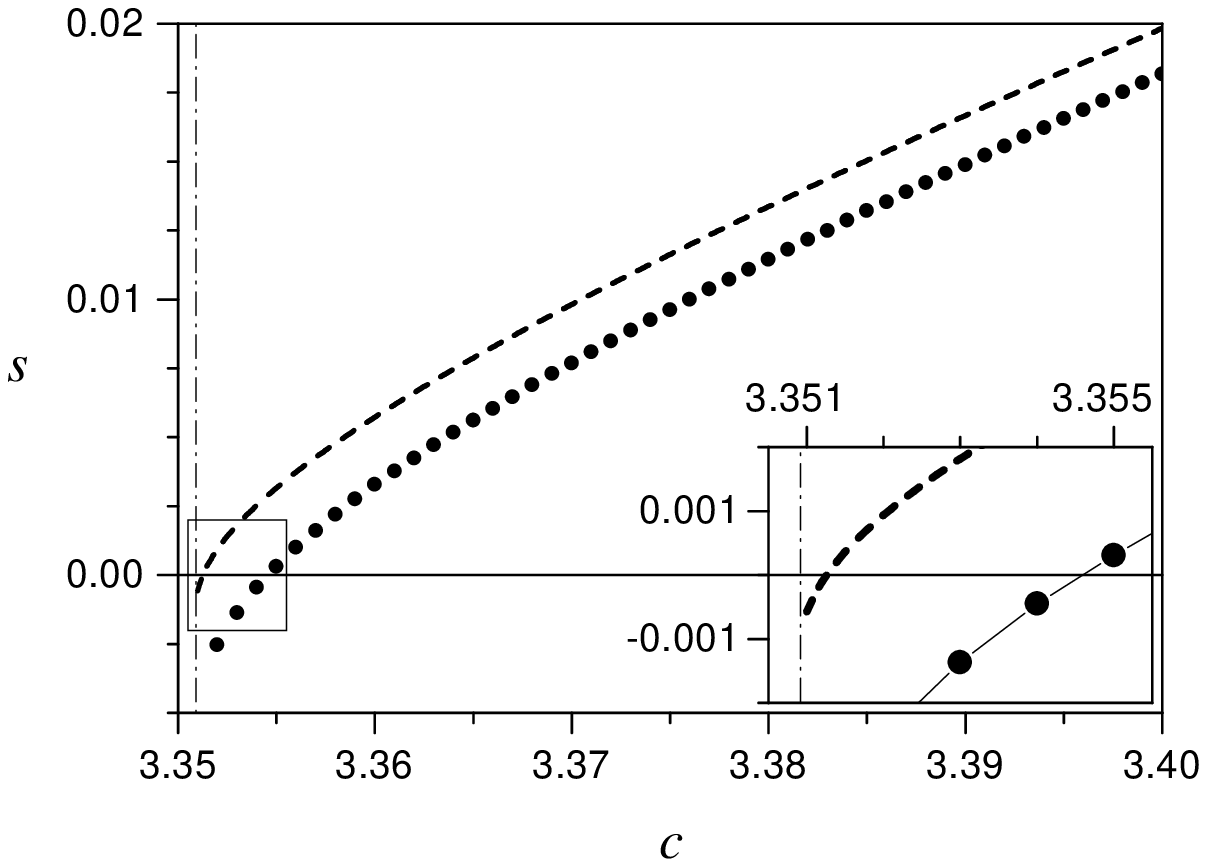}}
  \caption
  {
    Average subgraph size~$\phi$ (left)
    and entropy density~$s$ (right)
    as a function of the average vertex degree~$c$,
    for $3$-regular subgraphs (circles)
    and for the $3$-core (solid lines).
    Dashed lines denote results from the annealed approximation
    for $3$-regular subgraphs~\cite{PrettiWeigt},
    which provides an upper bound for the entropy, and a lower
    bound for the threshold.
    Thin dash-dotted lines mark the $3$-core threshold.
  }
  \label{fig:density-entropy-3}
\end{figure}
\begin{figure}
  \resizebox{70mm}{!}{\includegraphics*[35mm,155mm][165mm,245mm]{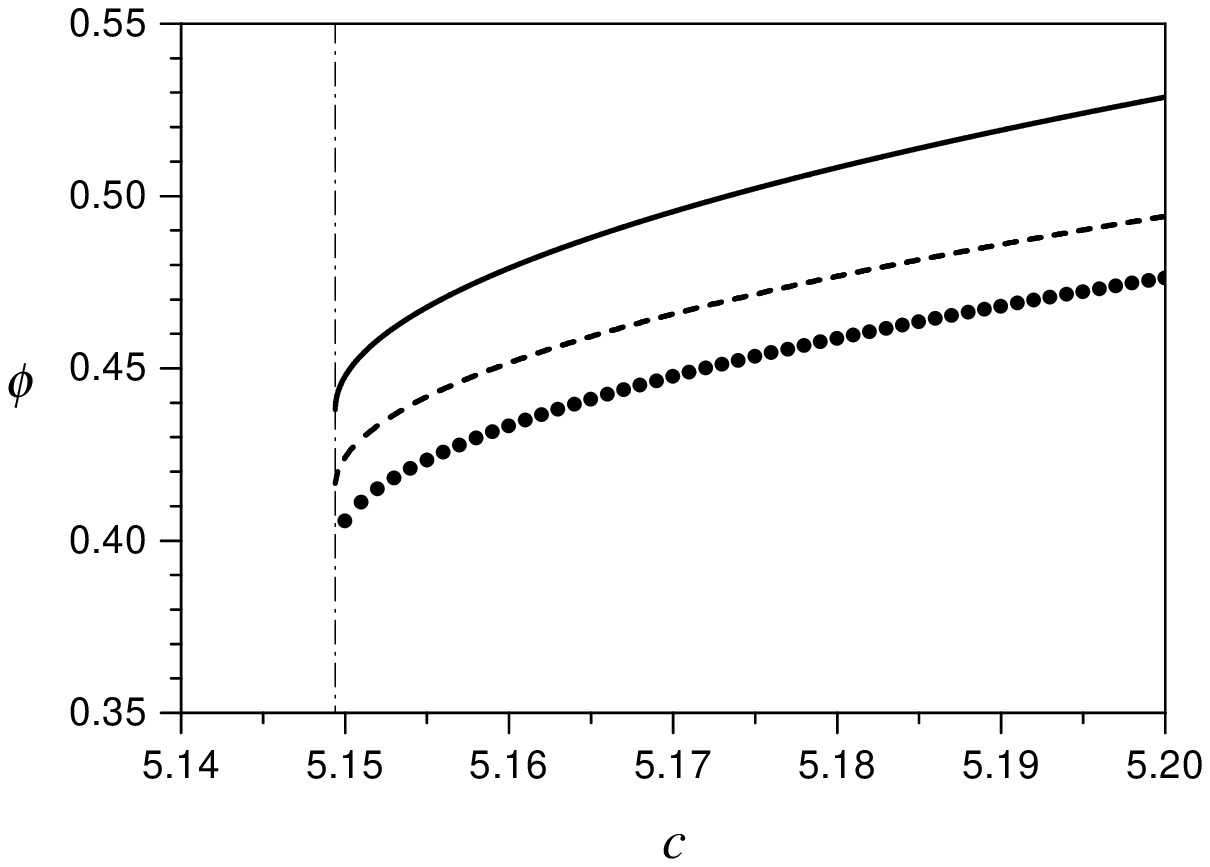}}
  \resizebox{70mm}{!}{\includegraphics*[35mm,155mm][165mm,245mm]{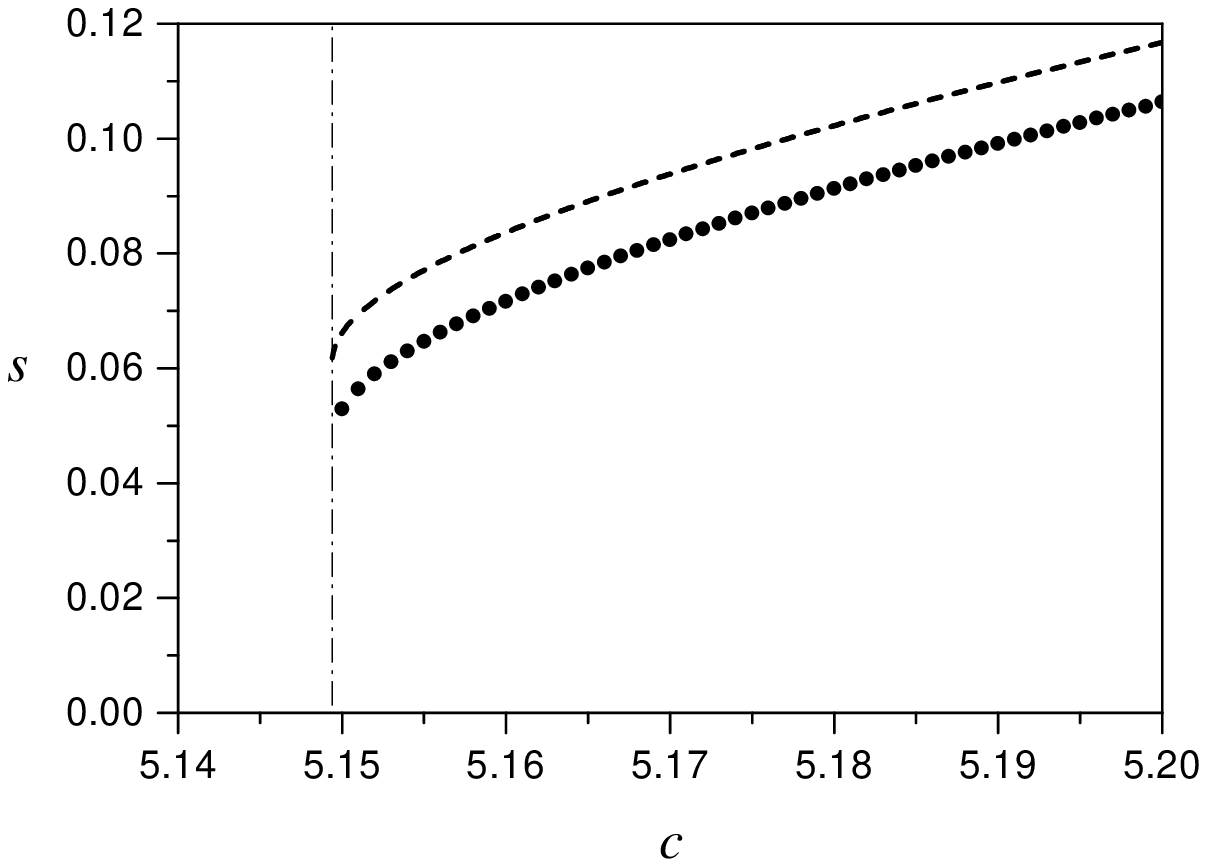}}
  \caption
  {
    The same as fig.~\ref{fig:density-entropy-3} for $q=4$.
  }
  \label{fig:density-entropy-4}
\end{figure}

In fig.~\ref{fig:density-entropy-3}, we report the results in
terms of average subgraph size (fraction~$\phi$ of vertices in the
subgraph) and quenched entropy density~$s$, for the case $q=3$. A
non-trivial solution appears at $c_\cor{3}$, but it has negative
entropy. The thermodynamically stable solution remains the
trivial, zero-entropy one, and no extensive $3$-regular subgraph
exists. Nevertheless, we can see that the entropy increases upon
growing average degree~$c$, and becomes positive at $c_\reg{3}
\simeq 3.3546$, where a {\it first-order phase transition} takes
place. The first $3$-regular subgraphs {\it appear
discontinuously}, containing immediately about 24\% of the full
graph. The situation is different for larger $q$ values. In
particular, we have investigated the cases $q=4,5$, for which one
can observe that the non-trivial solution still appears at
$c_\cor{q}$, but with an already positive entropy (see
fig.~\ref{fig:density-entropy-4}). We are thus led to conclude
that, contrary to the conjecture of Bollobas and
coworkers~\cite{BollobasKimVerstraete2006}, the emergence of
extensive $q$-regular subgraphs coincides with that of the
$q$-core, for $q>3$, whereas $q=3$ is a peculiar case.

\section{Stability of replica symmetry} So far, our results
rely on the assumption of replica symmetry. Does the inclusion of
possible replica-symmetry-broken solutions change the threshold
value? In order to investigate this issue, we have set up the
cavity equations for one-step replica symmetry breaking
(1-RSB)~\cite{MezardParisi2001}; details will be given
elsewhere~\cite{PrettiWeigt}. Using these equations, we have first
verified the local stability of the replica-symmetric solution
described above, i.e., that small perturbations decay
exponentially fast. Moreover, we have searched without success for
a non-trivial, locally stable 1-RSB solution, which could possibly
appear discontinuously. Based on these findings, we conjecture
that the replica-symmetric values $c_\reg{3} \simeq 3.3546$, and
$c_\reg{q}=c_\cor{q}$ for $q>3$, are exact. The mathematical proof
of these statements remains an interesting open problem. Let us
notice that, given the replica-symmetric character of the
transition, such a proof might be more easily accessible than the
proofs of exactness of statistical-physics results for threshold
phenomena in other constraint-satisfaction problems.

\section{Conclusion and outlook} In this letter, we have
analyzed the emergence (percolation) of $q$-regular subgraphs in
random graphs. Using the cavity method of statistical physics, we
have found that this happens in a first-order transition, i.e, the
subgraphs are immediately extensively large. These results are
based on a replica-symmetric calculation. Nevertheless, stability
with respect to replica symmetry breaking leads us to conjecture
that the observed threshold values are exact. For $q=3$, the
transition occurs in extreme vicinity to (but deviates from) the
well-known $3$-core percolation point $c_\cor{3} \simeq 3.3509$,
whereas, for larger $q$~values the threshold is found to coincide
with that of the $q$-core. Moreover, our method clarifies the
relationship between these apparently similar, but computationally
very different problems: Whenever $q$-regular subgraphs exist,
their union equals the $q$-core.

In a future publication~\cite{PrettiWeigt}, we shall further
elucidate the structure of such $q$-regular subgraphs. Our method,
as formulated here, allows to identify the entropy and thus also
the size of the {\it most frequent} $q$-regular subgraphs. One can
go beyond this by coupling the subgraph size to a conjugate
chemical potential, and analyze smallest and largest $q$-regular
subgraphs. Whereas it is mathematically clear that the $q$-core,
as the maximal subgraph of minimal degree at least~$q$, is unique,
a similar statement does not exist for maximal $q$-regular
subgraphs. A further interesting question is the one for
conditions of existence of a $q$-factor (i.e. a spanning
$q$-regular subgraph) in random graphs of given degree sequence.
According to a conjecture of Bollobas et al., these are expected
to exist if the minimal degree in the original graph is~$q+1$.

\acknowledgments We are indebted to Guilhem Semerjian for many
interesting discussions, and for communicating to us his results
prior to publication. We also acknowledge helpful discussions with
Michele Leone, Andrea Pagnani, and Lenka Zdeborov\'a.

\end{document}